# The onset and dynamics of avalanches in a rotating cylinder: From experimental data to a new geometric model

Christopher P. McLaren [1], Bernhard J. Leistner [1], Sebastian Pinzello [1], Eduardo Cano-Pleite [2], Christoph. R. Müller [1*]

[1] Department of Mechanical and Process Engineering, ETH Zurich, Leonhardstrasse 21, 8092 Zurich, Switzerland

[2] Thermal and Fluids Engineering Department, Carlos III University of Madrid, Avda. de la Universidad 30, 28911 Leganés, Spain.

[*] Corresponding author email address: muelchri@ethz.ch

### Abstract

Particle image velocimetry has been applied to measure particle velocities on the free surface of a bed of particles within a rotating cylinder during avalanching. The particle velocities were used to examine the validity of existing avalanche models and to propose an alternative model. The movement of particles depends on their location on the surface of the bed: particles located near the center of the bed travel the farthest, while the distance travelled decreases at an increasing rate for particles located farther from the center. The start of an avalanche can be determined to a single initiation point, that can also be located on the bottom half of the bed; the avalanche quickly propagates through the entire free surface, with 90% of the surface in motion within 257 ms. The experimental insight is used to formulate a new geometric model, in which three equal sized sections flow down the bed during an avalanche. The predictions of the model are confirmed by experimental mixing measurements.

## I. Introduction

Rotating drums are key for the processing of materials in the mineral, ceramic, metallurgical, chemical, pharmaceutical, waste and food industries [1–14]. A rotating cylinder can operate in different regimes. Among them, the avalanching regime is a quasi-periodic motion that is observed at low rotation speeds. In the avalanching regime the bed rotates as a rigid body until a section of the bed breaks off and moves down the free surface as an avalanche. The avalanche reduces the angle that the free surface of the bed forms with the horizontal. Fig 1. visualizes a simple geometric model of the bed surface before an avalanche (angle $\theta_s$) and after an avalanche (angle $\theta_t$). Following an avalanche, the bed returns to rigid body rotation and the angle between the bed's free surface and the horizontal increases until the next avalanche is initiated. Following the work of Bak *et al.* [15] of self-organized criticality, avalanching within rotating cylinders has received considerable attention [16–18].

Today there exist multiple competing models for the avalanching regime [19]. Metcalfe *et al.* [20] presented a model for particle mixing in the avalanche regime and compared predictions from their model with results from experiments performed in a pseudo two-dimensional rotating cylinder. The basis of their model was a geometric argument that the result of an avalanche event is to transport a wedge of material from the top half of the surface (light blue in FIG. 1**Error! Reference source not found.**) down the free surface such that after the avalanche it occupies a new wedge at the bottom half of the free surface (red in FIG. 1). The lines defining these wedges are the free surfaces of the bed before and after the avalanche has occurred and it was assumed that no particles outside of this wedge are affected by the avalanche. Based on this model it was argued that the problem of mixing within a system operating in the avalanching regime may be decomposed into two parts: the transport of wedges and mixing within wedges. In the absence of more detailed information, perfect mixing within the wedges was assumed. This wedge model was refined further by McCarthy *et al.* [21].

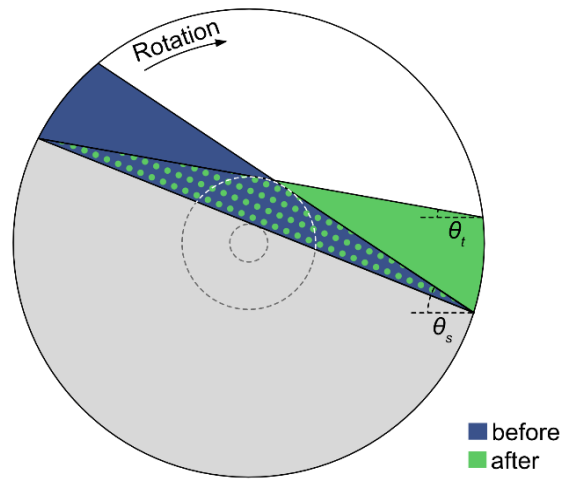

FIG. 1. Geometric models of an avalanche where $\theta_s$ is the angle between the bed surface and the horizontal before an avalanche and $\theta_t$ is the corresponding angle after an avalanche: the model proposed by Metcalfe *et al.* [20] has the purely blue wedge falling down to the purely green wedge. The model proposed by Lim *et al.* [22] starts width a wedge made of all areas containing blue falling down to the wedge of fully green plus green spotted.

Lim *et al.* [22] performed positron emission particle tracking (PEPT) studies of sand rotated in a horizontal cylinder that was operated in the avalanching regime. The PEPT data revealed that, on average, a particle on the free surface was involved in more than one avalanche before it rejoined the bulk of the bed in rigid body rotation. The average number of avalanches required for a particle to traverse the free surface was found to depend on the rotational

Froude number (Fr = $\omega^2 R/g$), where $\omega$ is the rotational speed, $R$ is the radius of the cylinder and $g$ is the gravitational acceleration. At low values of Fr, particles remained on the free surface for an average of 4.25 avalanches, whereas this number drops to one avalanche when the bed was operated in the rolling regime with Fr = $1.47\times10^{-4}$. They found the average number of avalanches required for a particle to traverse the free surface to linearly decrease with increasing Fr. These observations do not support the geometric wedge model of avalanches proposed by Metcalfe *et al.* [20] since the geometric model implies that particles traverse the free surface in a single avalanche. In light of their observations, Lim *et al.* [22] proposed a modified wedge model in which the centroid of the avalanching material moves a distance 1/3 of the chord length down the free surface, where *l* is the length of the free surface (see FIG. 1).

So far, studies that have investigated mixing in the avalanching regime in rotating cylinders have not tracked the overall motion of all particles. Lim *et al.* [22] tracked only a single particle using PEPT. Du Pont *et al.* [23] tracked the flowing surface averaging the velocities along the surface flow. Kiesgen de Richter *et al.* [24] studied grain rearrangements by using a camera and observing the acoustic amplitude across the bed with using a piezo transducer and receiver, allowing them to observe changes in the weak contacts, i.e. contacts with less than the designated cutoff force. Zaitsev *et al.* [25] observed a slow relaxation of particle displacements to an equilibrium position after the avalanche. Both Kiesgen de Richter *et al.* [24] and Zaitsev *et al.* [25] observed precursors to avalanches, where small changes in the packing lead to instabilities that flow of the surface did not occur.

In order to examine the validity of the geometric models proposed by Metcalfe *et al.* [20] and Lim *et al.* [22], Particle Image Velocimetry (PIV) is applied in this work to investigate the motion of the particles on the bed surface during avalanches in a horizontal rotating cylinder. Utilizing our experimental findings, an alternative geometric model for mixing during avalanching is proposed and compared against experimental measurements.

## II. Experimental Setup

In this work three drums of different diameters, see TABLE I, were studied. The drums were constructed out of clear Perspex, allowing two cameras an unobstructed view to the top surface and the side of the drums. Drum 2 was filled with particles to fill levels ranging from 20 % to 70 % with 5 % increments. Three differently sized spherical particles, as shown in TABLE II were studied. In order to perform PIV, 10% of the particles though otherwise identical, were darker colored, to allow tracking during the PIV image correlation. The cylinder was driven by a motor (Maxon EPOS 2 24/5) and the free surface of the bed was imaged using a high-speed camera (Nikon, 496RC2) with a frame rate of 100 fps and a resolution of 1280x1024 pixels for PIV purposes. A Canon EOS camera filmed the side of the drum, to determine the slope of the free surface of the bed.

The particle velocities were calculated using the MATPIV 1.6.1. software [26]. The size of the PIV interrogation window was iteratively reduced from 64x64 to 32x32 pixels, using four iterations. Once the velocities were determined, a signal-to-noise filter (s/n = 1.3) and global and local mean filters were applied to detect outliers [27]. To ensure statistical significance of the results, 14-20 avalanches were averaged. The onset of avalanching was defined as the time when the first particle assembly (PIV window) exceeded a threshold velocity of 3 mm/s. This threshold velocity was found to have a negligible impact when set between 1mm/s and 20 mm/s.

TABLE I. Drums

|  | Drum 1 | Drum 2 | Drum 3 |
|---|---|---|---|
| Inner diameter $D$ [mm] | 94 | 140 | 290 |
| Length [mm] | 400 | 400 | 400 |

TABLE II. Spherical particles

|  | Ballotini 1 | Ballotini 2 | Ballotini 3 |
|---|---|---|---|
| Diameter $d$ [m] | 1.0 – 1.3 | 2.0 – 2.4 | 2.85 – 3.45 |
| Material | glass | glass | glass |
| Density [kg/dm³] | 1.49 | 1.45 | 1.43 |

## III. Results

FIG. 2 shows the time evolution of the slope of the free surface of the bed for drum diameters of 94 mm, 140 mm, and 290 mm. For all drum diameters the evolution of the slope of the free surface shows a "sawtooth" pattern that is characteristic of the avalanching regime; individual avalanches can be identified as a rapid decrease in the slope of the free surface and between avalanches the slope increases linearly with time, indicating rigid body rotation.

It can be observed that avalanching is a quasi-periodic phenomenon with fluctuations in both the magnitude of the slope change during an avalanche and the time between successive avalanches. In agreement with Daerr *et al.* [28] the avalanches appear to become more regular as the drum diameter is increased. This was also observed by Daerr *et al.* [28] when increasing the rational speed, i.e., the Froude number.

In the following, the particle motion at the start of an avalanche will be considered. Rearrangements on the surface as precursors described in previous studies were observed but are outside of the scope of this of this study [24].

In this work the start of motion was defined as the moment when a window of particles exceeded the threshold velocity $v_t$ of 3 mm/s. The avalanche's initiation time was found to be insensitive to this velocity threshold. When varying the threshold between 1 and 20 mm/s, the initiation time is delayed at most by 20 ms, which is less than 2 % of the duration of an avalanche. Furthermore, the location of the initiation point of the avalanche is unaffected by the variation of the velocity threshold and the propagation of the avalanche at the free surface proceeds at the same rate for thresholds between 1 and 20 mm/s. FIG. 3 shows the time after the initiation of an avalanche, i.e. the elapsed time between the beginning of the avalanche until a specific location of the free surface exceeds $v_{L}$. Data are shown from three avalanches for a drum diameter of 140 mm, a particle size range of diameter 1.0 - 1.3 mm and fill level of 25 %. It can be seen that an avalanche can initiate at a variety of locations on the free surface. The fraction of the surface in motion increases linearly for roughly 10 % of the duration of the avalanche and then the propagation slows down. On average 90 % of the free surface exceeds $v_{L}$ within 257 ms ± 98 ms of the start of the avalanche. The average duration of the avalanche is 1.5 seconds, thus roughly the first 1/5 of the avalanche duration is needed to put the almost entire bed surface in motion. These observations are confirmed in FIG. 4, which shows the fraction of the free surface that is in motion as a function of time. The fraction of the fee

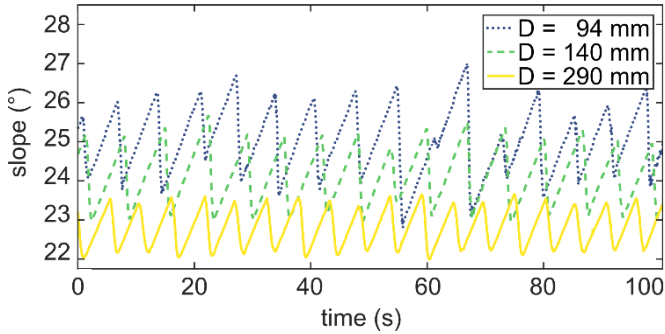

FIG. 2. The slope of the free surface as a function of time for three drum diameters: 94 mm (dotted blue), 140 mm (dashed red) and 290 mm (yellow) rotating at 0.066 rpm and a fill level of 25 %. The particles are ballotini with a diameter of 1.0 - 1.3 mm.

FIG. 5 shows velocities of the avalanching material at the start of an avalanche, during an avalanche, and at the end of an avalanche. Data are shown for an avalanche with a drum diameter of 140 mm, a particle size range of diameter 1.0 - 1.3 mm and fill level of 25 %. At the start of the avalanche the first movers begin their trajectory down the slope while the other particles do not show any relative motion. The maximal speed averaged across the bed surface (24 mm/s) was attained 0.59 seconds after the start of the avalanche (around 44 % of the duration of the avalanche). At this point the entire bed is in motion and appears to move uniformly down the slope as shown in the middle panel of FIG. 5. In contrast to the beginning of the avalanche, where the trigger of the avalanche could be traced to an individual starting point, the end of the avalanche has a slow random relaxation of particle displacements to an equilibrium as observed by Zaitsev *et al.* and De Boeuf *et al.* [25,29].

The high temporal and spatial resolutions of the PIV measurements reported here allow to estimate the trajectories of particles during an avalanche by integrating the PIV data. Note that PIV calculates the displacement of patterns, rather than of individual particles, with the result that the trajectories obtained in this manner do not represent the exact paths of individual particles. Here the term "pseudo particle" will be used to denote a pattern whose trajectory has been calculated by integrating PIV data.

FIG. 6 shows the distances moved by pseudo particles during an avalanche as a function of the starting position of the pseudo particle. Data are shown for an avalanche in a drum of diameter 140 mm and fill level of 25 %. The particles are glass ballotini with a diameter of 1.0 - 1.3 mm.

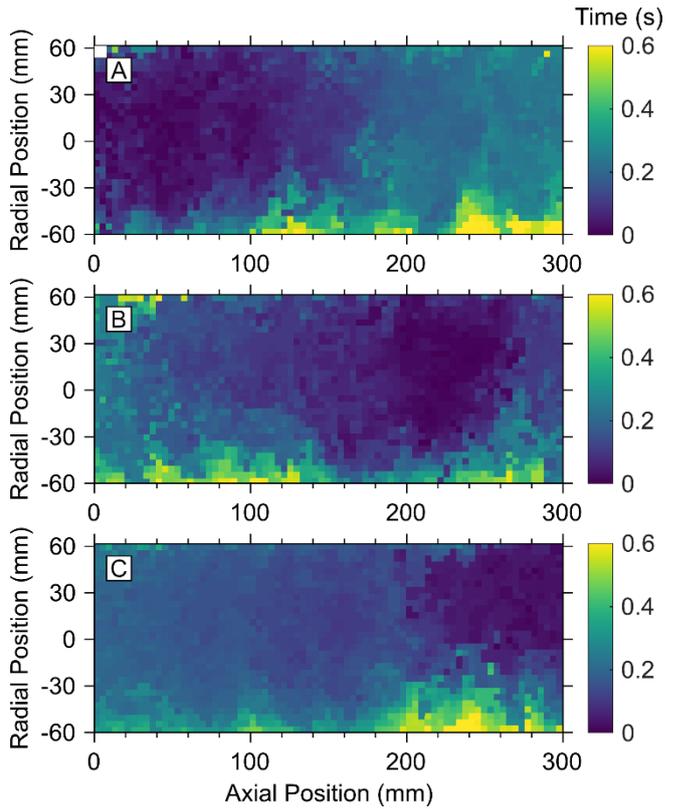

FIG. 3. Local avalanche initiation time shown for three independent avalanche events. The color indicates the time after the start of the avalanche for which the velocity at a point on the free surface exceeds 3 mm/s. The cylinder has an internal diameter of 140 mm and rotates at 0.066 rpm. The fill level is 25 % and the particles are glass ballotini with a diameter 1.0 - 1.3 mm.

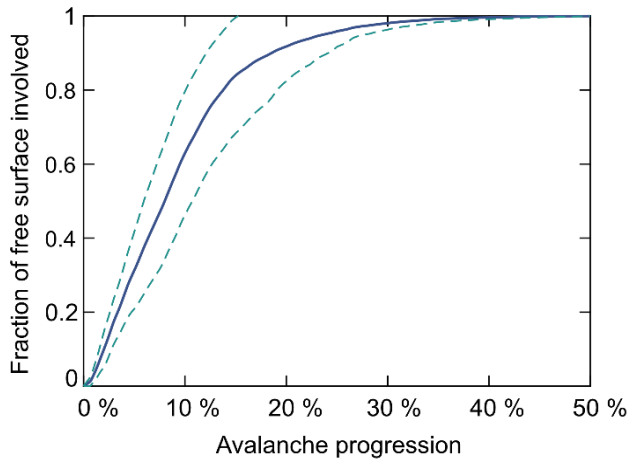

FIG. 4. The fraction of the bed surface involved in an avalanching event as a function of the fraction of the duration of an avalanche event. The mean and standard deviation were calculated from the data of 20 avalanches. The cylinder has a diameter of 140 mm and rotates at 0.066 rpm. The fill level is 25 % and the particles are glass ballotini have a diameter of 1.0 - 1.3 mm.

Distances travelled for particles located in the lowest 20% of the chord are not shown as they could have been overtaken by particles starting higher on the surface of the bed. PIV is only able to track the velocities of particles that remain on the surface. Thus, when a pseudo particle is overtaken by material flowing over, it is no longer trackable. Any subsequent velocity for this pseudo particle would correspond to the pseudo particle that had overtaken it. To avoid using the velocities of a particle that has overtaken the original particle, a control was set in place. The initial positions of pseudo particles were separated by the size of a PIV interrogation window to the pseudo particle below. In this work a pseudo particle was assumed to have been overtaken if a particle that started higher on the surface of the bed approached to a distance of less than half an interrogation window. It was found that only pseudo particles in the lowest 20 % of the chord could be overtaken.

FIG. 7 shows the average distances travelled by particles down the free surface during an avalanche for a series of 16 avalanches. As with FIG. 6, the distance travelled for particles starting in the lowest 20 % of the chord are not shown.

The maximum distance travelled is for the particles starting in the center of the bed surface, travelling on average 20 % of the chord length. The farther the particle from the center, the distance travelled by the particle decreases at an increasing rate. Due to the low Froude number, particles here travel less than 1/3 of the chord length as hypothesized in the model of Lim et al. [22] Above 85 % of the chord length from the bottom, there is a dramatic decrease in the distance travelled, with the top 5% of the bed surface only travelling half as far, i.e. 10% of the chord length.

As shown in FIG. 8, the maxima of the mean speed averaged over the entire surface as well as the average distance travelled increase with growing drum diameters. Similar to FIG. 6, the distance travelled for particles starting in the lowest 20 % of the chord were again not taken into account. When increasing the Froude number by increasing the rotational speed Daerr *et al.* [28] observed an increase in flow speed, which is in accordance with our findings.

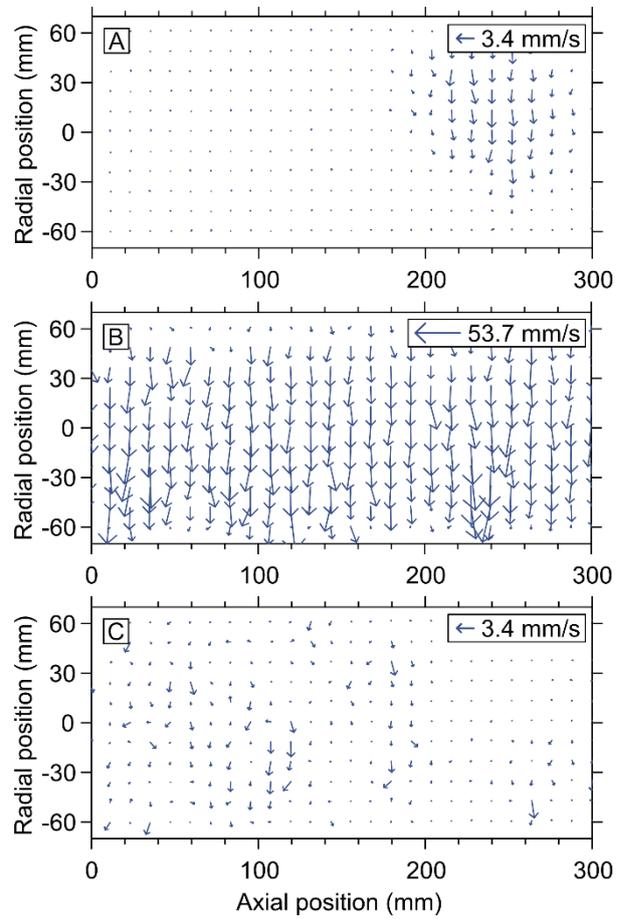

FIG. 5. Velocity of particles at the bed surface during an avalanching event: (a) at the start, (b) at the maximal speed and (c) at the end of an avalanche, based on PIV calculations. The beginning of an avalanche is defined as the point in time when the first particles exceeded a velocity of 3 mm/s. The maximal speed attained by the avalanching material in the avalanche shown was 54 mm/s. The end of the avalanche was defined as the point when the speed of all pseudo particles was below 3 mm/s. The cylinder has an internal diameter of 140 mm and rotates at 0.066 rpm. The fill level is approximately 25 % and the particles are glass ballotini with a particle size range 1.0 - 1.3 mm.

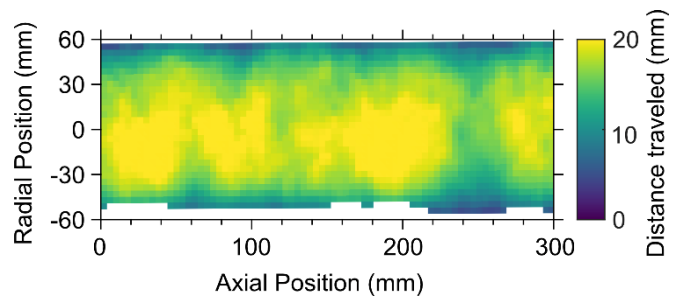

FIG. 6. Distance travelled as a function of the starting position for an inner drum diameter of 140 mm. The particles used were ballotini with a diameter of 1.0 - 1.3 mm. The drum rotated at 0.066 rpm and was filled to a fill level of 25 %.

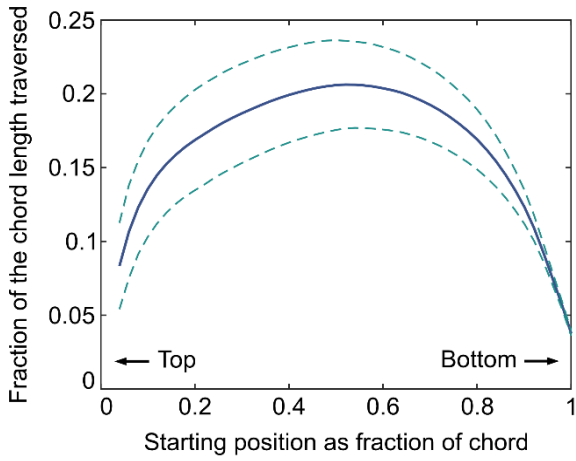

FIG. 7. Mean fraction of the chord length travelled as a function of the starting position along the chord. The data has been averaged from 16 avalanches and the dashed lines show one standard deviation. The cylinder has a diameter of 300 mm and rotates at 0.066 rpm. The fill level is 25 % and the particles are glass ballotini with a diameter of 1.0 - 1.3 mm.

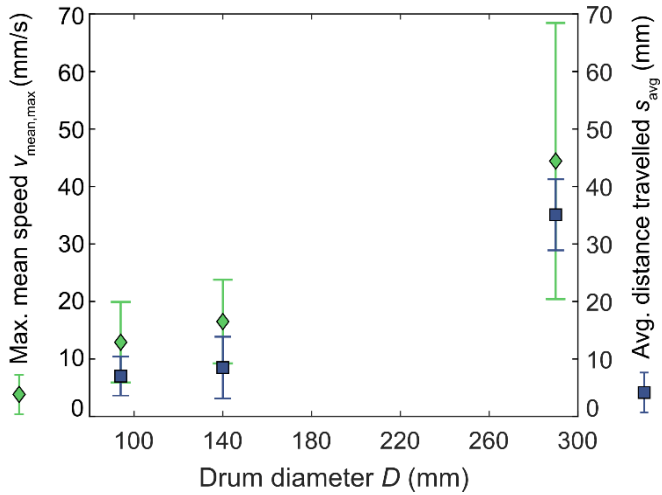

FIG. 8. Maxima of the mean speed of the entire bed surface (green) and the average distance travelled (blue) averaged over the entire surface with their respective standard deviations as a function of the drum diameter. The drums rotated at 0.066 rpm. The particles used were ballotini with a diameter 1.0 - 1.3 mm and a fill level of 25 %. The lowest 20 % of the chord was not accounted for to calculate the distance travelled, as particles located in this part of the chord may have been overtaken by particle initially located higher up the chord.

## IV. Discussion

The distance travelled by pseudo particles along the free surface of the bed depends on their starting location. As seen in FIG. 7, the particles that start in the center of the chord of the free surface travel the farthest and the farther the particle are from the center, the distance travelled by the particle decreases at an increasing rate. This behavior can be understood based on the work by Daerr *et al.* [28], which showed that the movement of particles in an avalanche is driven by two effects: the loss of support as the lower particles slide away, and an increase in weight due to downward motion of particles that were initially located higher up. At the highest point of the free surface the weight is minimal, while at the lowest point of the free surface the motion of particles is hindered by the cylinder wall. On the other hand, at the center of the free surface both effects occur, leading in turn to high avalanche velocities (FIG. 5b).

Metcalfe *et al.* [20] proposed a geometric model for avalanches in which it was assumed that an avalanche transports a "wedge" of material from the upper half of the free surface to the lower half of the free surface as shown in FIG. 1. In this model the lower half of the free surface is not involved in the avalanche and is simply covered by material flowing from the upper half of the free surface. The data presented here is not consistent with this simple geometric model. FIG. 3 and FIG. 5 clearly demonstrate that avalanches are being initiated in the center of the free surface (FIG. 3). The avalanche is then propagated through the surface of the bed, setting its entire surface in motion (FIG. 5), including the lower half of the bed.

Lim *et al.* [22] proposed a modified wedge model. The model of Lim *et al.* [22] assumes that the avalanching material moves an average distance of 1/3 of the chord down the free surface, rather than travelling the entire free surface in one avalanche. Their defined wedge is made up of two chords: the top surface of the bed before and after the avalanche. In accordance with Lim *et al.* [22], who suggested that the number of avalanches required for a particle to traverse the free surface to decrease linearly with increasing Fr, we found the average distance travelled by particles in one avalanche to increase with increasing drum diameter. In FIG. 8 a linear correlation is possible, as would be expected according to the relation from Lim *et al.* [22], between the drum diameter and the average distance travelled In Lim *et al.*'s [22] model due to the shape of the wedge, which is shown schematically in FIG. 1, most of the material which descends to the lower wedge is initially located at the top of the free surface. However, in our experiments we observe that particles located at the very top of the bed move significantly less during an avalanche than those in the center of the free surface. Therefore, although the data reported here suggest the model proposed by Lim *et al.* [22] is unlikely, we cannot disprove this model based on these data alone.

Instead, based on our experimental results we propose a new geometric avalanche model. The new model attempts maintain a simple geometry just as the models of Metcalfe *et al.* [20] and Lim *et al.* [22], yet take into account the PIV results which show that the particles in the center of the free surface move the farthest. The proposed model is sketched in

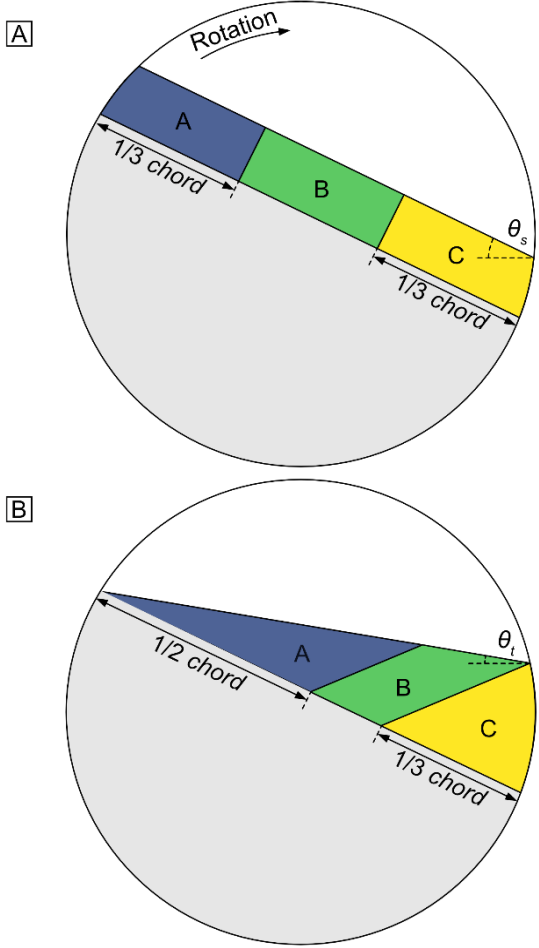

FIG. 9. Here, the flowing (avalanching) layer is split into three equal "boxes" which deform into a wedge shape over an avalanching event. This model reflects better the PIV results in that on average particles in the middle "box" travel farther than particles in the bottom "box" and the same distance as particles in the top "box." Hence, although the middle section still does not move the farthest it is an improvement on both the Metcalfe *et al.* [20] and Lim *et al.* models for which the top section travel the furthest down the free surface.

In order to compare the accuracy and visualize the difference of these three models, we recorded the mixing of an initially segregated bed from the lateral face of the cylinder. This methodology has been used previously by Metcalfe el. al. [20] to assess the accuracy of their model. For all models, we assumed perfect mixing within the segment of the free surface, although Lim *et al.* do not state explicitly this assumption. (Metcalfe *et al.* [20] state explicitly that perfect mixing in a segment is assumed.) First, we attempted to compare the predictions of the different models with the experimental data by computing a global parameter, i.e. the center of mass of the pink particles. Figure S1 compares the position of the center of mass of the pink particles (fill level 80 %, drum diameter 200 mm, particle size range 1.0-1.3 mm) predicted by the three models with the experimental measurements. In the beginning the models and the experimental data show very good agreement, while towards longer rotation times, larger differences between the individual models and the experimental data become apparent. Towards the end of the mixing experiments, it appears as the new model proposed here is closest to the experimental data, yet a clear assessment is difficult.

In a second step, we used key transition points for comparison between the models and the experimental results. The first key point is after the first avalanche. A side view of the bed after the first avalanche in the simulations and experiment is provided in FIG. 10(a,b). In the experiment we observe a pink wedge, tapered to the outside of the cylinder, after the first avalanche. In the Metcalfe *et al.* [20] model a wedge has been formed but it tapers in towards the center of the cylinder, which does not agree with the experimental observations. The wedge predicted by the model of Lim *et al.* [22] also tapers in the opposite direction to the highest point of the free surface to what is observed experimentally. Owing to the particular construction of the wedges in the Lim *et al.* [22] model and with the assumption of perfect mixing, some black particles are predicted in the upper half of the free surface after the first avalanche. As a matter of fact, Lim et al. [22] do not describe how mixing occurs in the wedge and it is only mentioned that particles move on average 1/3 of the length of the chord. On the one hand, if perfect mixing is taking place inside the wedges and its centroid moves *1/3* of the chord length, as considered in FIG. 10, an unrealistic behavior of the black particles would take place. On the other hand, if all particles on the surface of the bed move on average 1/3 of the chord length, the whole wedge would slide down, thus not making possible to have the wedge tapering towards the top they predict.

The new model proposed here overcomes the abovementioned limitations of both Metcalfe *et al.* [20] and Lim et. al [22] models and predicts a physically realistic formation of a wedge of pink particles that is tapering in the same direction as the experiments, i.e. towards the lower edge of the free surface.

Another further key transition moment is when black particles start to avalanche down over the pink particles. This point in the rotation cycle allows for another comparison between the three models and experiments (FIG. 10(c,d)). A red line along the perimeter of the cylinder of equal length allows for an easier comparison between the experiments and the different models. In the models of Lim *et al.* [22] and Metcalfe *et al.* [20], the black beads close to the perimeter have travelled farther down compared to the newly proposed model and the experiments. A further region that allows for a quantitative comparison is the size of the core, i.e. the unmixed center part of the cylinder that exists for fill levels greater than 50 % [20]. The distance from the top of the free surface to the unmixed core, $h_c$, is marked in FIG. 10d and is used for comparison. We observe that $h_c$ as determined in the experiments is larger than the value predicted in any of the models. Nonetheless, the model proposed in this work gives values of $h_c$ that are the closest to the experimental value. TABLE III ($h_c$ for 60 % and 80 % fill levels) confirms quantitatively the improved accuracy of the newly proposed model, albeit there still exist some appreciable differences to the experimental results.

TABLE III. Distance from the surface of bed to the unmixed core of the bed ($h_c$). The unmixed core is the part of the bed that is not involved during an avalanche and which therefore remains unchanged throughout all rotations

| | $h_c$ 60 % fill [mm] | $h_c$ 80 % fill [mm] |
|---|---|---|
| Experiment | $10.1 \pm 2.0$ | $6.6 \pm 1.3$ |
| New Model | $4.0 \pm 0.3$ | $2.5 \pm 0.5$ |
| Lim *et al.* | $0.9 \pm 0.5$ | $1.3 \pm 0.4$ |
| Metcalfe *et al.* | $0.1 \pm 0.2$ | $0.2 \pm 0.3$ |

FIG. 9. The new geometric avalanche model proposed here based on the experimental PIV data aquired. The flowing layer of the bed is separated into sections A, B and C: (a) The flowing (avanlanching) layer before the avalanche event is initiated; (b) Deformation of the initial segments during the avalanching event considering that particles located in the center of the bed move the farthest during an avalanching event.

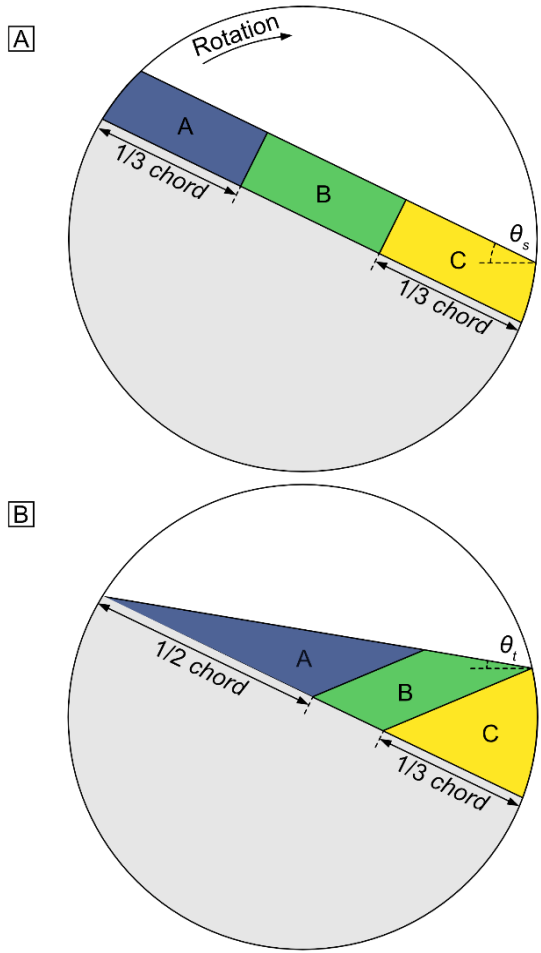

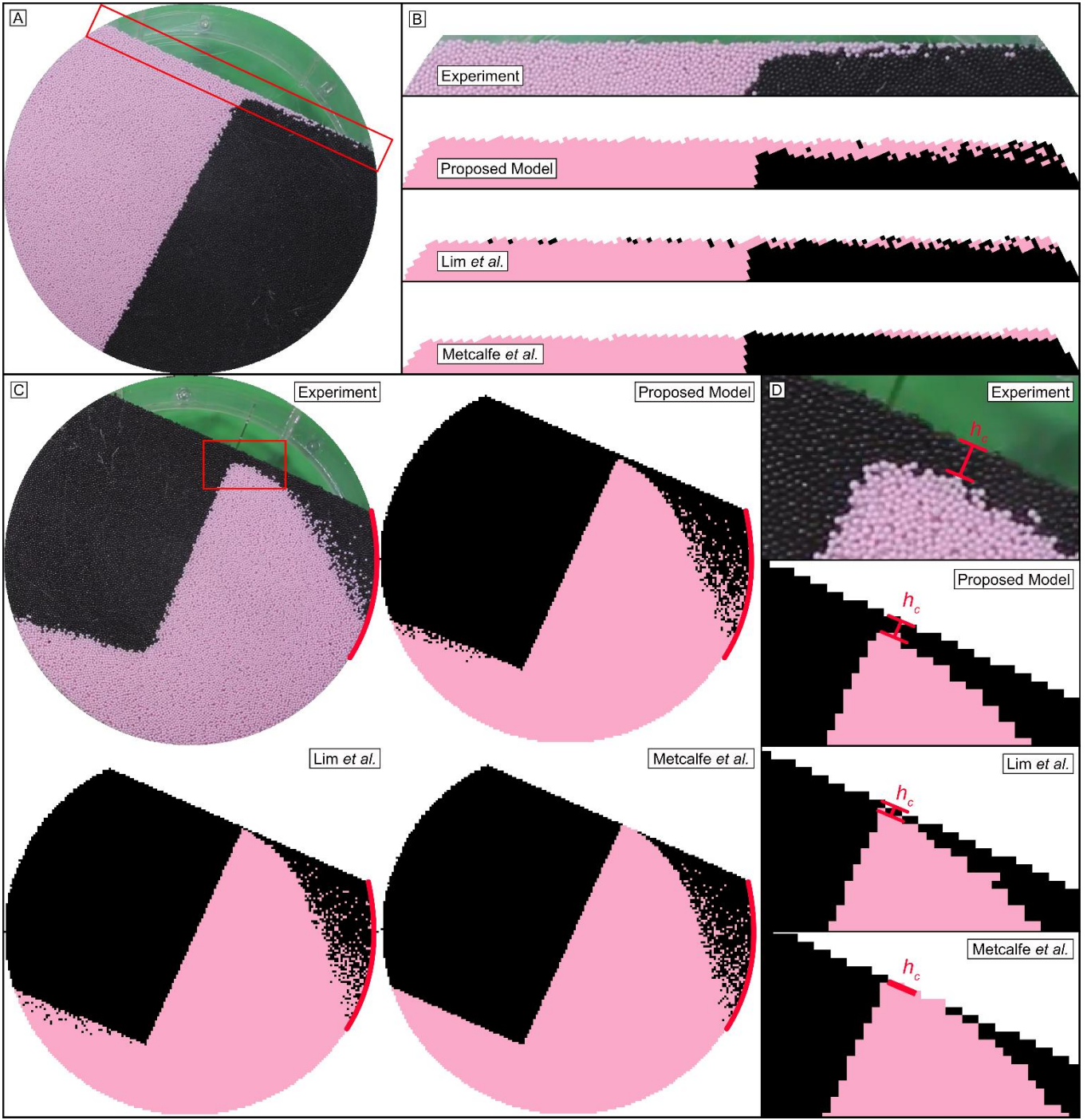

FIG. 10. Comparison between the predictions of the models of Metcalfe *et al.* [20] and Lim *et al.* [22], the newly proposed model and experimental data. The red rectangle (a) show the area that has been zoomed into in (b). The snapshot in (a) gives a side of the entire bed after one avalanche. The snapshots in (b) zoom into the top 20 mm of the experimental bed and the predictions of the models of Metcalfe *et al.* [20] and Lim *et al.* [22] and the newly poposed model. (c) Snapshots of the beds after a rotation by 204°. A red line spanning 59° is added to aid a visual comparison between the predictions of the model and the experimental data; (d) zoom into a region spanning from the top surface to the start of the unmixed core of the bed.

## V. Conclusion

PIV has been performed to examine the movement of particles along the free surface of an avalanche in a rotating cylinder. The angle of repose of the bed was also tracked.

A "sawtooth" pattern can be seen for the slope of the bed as a function of time. The individual avalanches can be observed by a rapid decrease in the slope angle, with each avalanche followed by a linear increase in slope angle during rigid body rotation. As the diameter of the cylinder, and hence the Froude number, is increased, the avalanches become more regular, as also observed by Daerr *et al.* [28].

Avalanches appear to begin from an initiation point anywhere on the free surface and propagate quickly such that more than 90 % of the bed is in motion by about 20 % of the duration of an average avalanche from its start. At the point when the maximum speed in an avalanche was attained, the entire bed is uni-directionally travelling down the chord. The particles still in movement at the end of an avalanche appear to be randomly scattered on the free surface with multi-directional velocities similar to the results of Zaitsev *et al.* and De Boeuf *et al.* [25,29]. The distances travelled along the free surface of the particle bed are a function of their starting location along the chord of the free surface. The particles that travel the farthest are located at the center of the free surface. Farther from the center, the distance travelled by the particles decreased at an increasing rate. Based upon these results an updated geometric model was proposed, where the flowing surface is divided into three sections which deform into a wedge during an avalanching event. The newly proposed model shows a better agreement with experimental mixing data when compared to the models of Metcalfe *et al.* [20] and Lim *et al.* [22], albeit the improvement is only visible in a few specific locations.

## Acknowledgement

We acknowledge the Swiss National Science Foundation (grant number SNF_182692) for financial support and James Third for valuable input in the beginning of the project. This publication was created as part of NCCR Catalysis, a National Centre of Competence in Research funded by the Swiss National Science Foundation. Eduardo Cano-Pleite acknowledges support from the CONEX-Plus program funded by Universidad Carlos III de Madrid and the European Union's Horizon 2020 program under the Marie Sklodowska-Curie grant agreement No. 801538.

## Supplementary Info

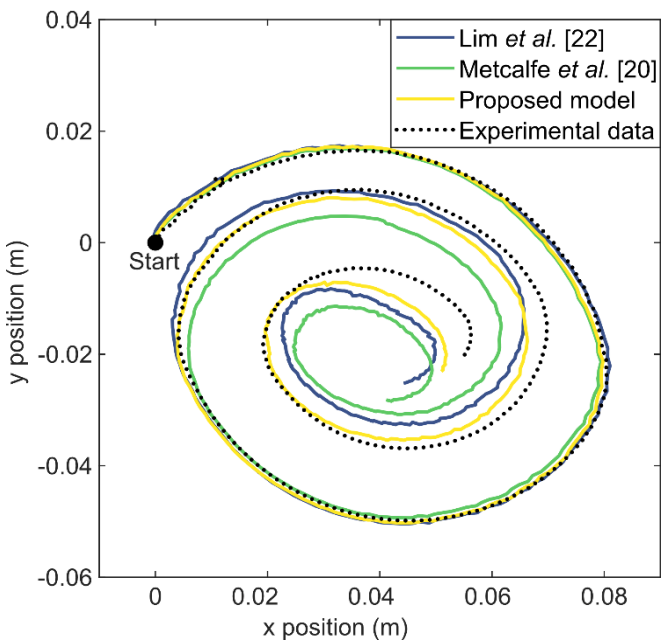

FIG. SI 1. The pink particles' change in center of mass during two drum rotations is shown. We compare our proposed model and experimental data with the models of Lim et al. [22] and Metcalfe et al. [20]. The positions are given relative to the starting point.